# Semimicroscopic description of isoscalar giant multipole resonances in medium-mass closed-shell nuclei

M.L. Gorelik [1], S. Shlomo [2], B.A. Tulupov [3], and M.H. Urin [4]

[1] *Moscow Economic School, Moscow 123022, Russia*

[2] *Cyclotron Institute, Texas A&M University, College Station, TX 77843, USA*

[3] *Institute for Nuclear Research, RAS, Moscow 117312, Russia*

[4] *National Research Nuclear University "MEPhI", Moscow 115409, Russia*

Main properties of isoscalar giant multipole resonances (up to $L = 3$), including $L = 0$, and 2 overtones, in medium-mass closed-shell nuclei are described within the semimicroscopic particle-hole dispersive optical model. Main properties are characterized by the energy-averaged strength distribution, projected (one-body) transition density, and probabilities of direct one-nucleon decay. Calculation results obtained for characteristics of the mentioned resonances in the $^{48}$Ca, $^{90}$Zr, and $^{132}$Sn nuclei are compared with available experimental data.

## I. INTRODUCTION

Being a microscopically based extension of the standard [1,2] and nonstandard [3] versions of the continuum-random-phase approximation (cRPA) to taking the spreading effect into account, the particle-hole dispersive optical model (PHDOM) has been formulated [4] and implemented to describing main properties of various giant resonances (GRs) in medium-heavy closed-shell nuclei [5-10]. The GR main characteristics, which are determined for a wide excitation-energy interval, include the energy-averaged strength distribution and projected (one-body) transition density, both related to an appropriate probing operator, and the probabilities of direct one-nucleon decay. PHDOM can be used for describing these characteristics, because within the model, the main relaxation modes of high-energy (p-h)-type states, associated with GRs, are together taken into account. These modes, Landau damping and the coupling of mentioned states to the single-particle (s-p) continuum are described microscopically (in terms of a mean field and p-h interaction), while the coupling to many-quasiparticle configurations (the spreading effect) is treated phenomenologically (in terms of the properly parameterized energy-averaged p-h self-energy term). The imaginary part of the self-energy term determines the real part via a proper dispersive relationship. Landau-Migdal p-h interaction and a realistic (Woods-Saxon-type) phenomenological mean field consistent with the spin-less part of this interaction are used as input quantities in implementations of PHDOM. The self-consistency conditions appear due to isospin symmetry and translation invariance of the model Hamiltonian. Parameters of the mean field and p-h interaction are found from independent data, while the "spreading" parameters (they determine the strength of the imaginary part of the energy-averaged p-h self-energy term) are considered as adjusted quantities.

In Ref. [10], the PHDOM current version has been adopted to describing main characteristics of isoscalar giant multipole resonances (ISGMPRs) in medium-heavy closed-shell nuclei. The $L = 0 - 3$ resonances together with $L = 0, 2$ overtones have been considered. A cRPA-based description of low-energy isoscalar collective states, including the $1^-$ spurious state related to center-of-mass motion, has been also taken into consideration. The adopted model has been implemented to describe main characteristics of the mentioned resonances in the $^{208}$Pb nucleus, taken as the heaviest doubly-closed-shell nucleus. A rather reasonable description of respective experimental data has been obtained. Some of results obtained within PHDOM for $L = 0 - 3$ ISGMPRs were compared with the respective results of calculations obtained within the microscopic RPA-based approach of self-consistent Hartree-Fock (HF), using Skyrme-type forces [2,11].

The studies of Ref. [5-10] complete the "testing" stage of PHDOM-based implementations to the description of various GRs mainly in $^{208}$Pb. The next stage consists in systematic evaluating, within PHDOM, the main characteristics of families of isoscalar and isovector GRs in medium heavy closed shell parent nuclei. The first step in this direction has been recently done in applying to the Gamow-Teller and charge exchange spin-monopole GRs [12].

The present work is a direct continuation of the study undertaken in Ref. [10], Using the same "spreading parameters", we extend this study to evaluation of main characteristics of the above-mentioned six ISGMPRs in medium-mass closed-shell nuclei $^{48}$Ca, $^{90}$Zr, and $^{132}$Sn. For reader's convenience, we show the model relations, which are directly used in calculations (Section II). Choice of model parameters, calculation results and discussion of the results are given in Section III. Section IV contains our concluding remarks.

## II. MODEL RELATIONS

Being related to a description of the GR strength distribution and double transition density, PHDOM has been originally formulated in terms of the energy-averaged p-h Green function (effective p-h propagator), which obeys the Bethe-Goldstone-type equation [4]. The non-homogeneous term in this equation (the "free" p-h propagator) is the PHDOM key quantity, in which Landau damping, the s-p continuum, and the spreading effect are together taken into account. The "free" p-h propagator is related to the model of non-interacting and independently damping p-h excitations. In the above-mentioned equation, Landau-Migdal forces are used, as the p-h interaction responsible for long-range correlations. For describing non-spin-flip GRs, the following part of these forces is used:

$$F(x_1, x_2) \rightarrow (F(r_1) + F'\vec{\tau}_1\vec{\tau}_2)\delta(\vec{r}_1 - \vec{r}_2). \tag{1}$$

To evaluate the double transition density depended only on nuclear structure, it is necessary to solve the Bethe-Goldstone equation for the effective p-h propagator. However, for a practical use, the specific (projected) one-body transition density can be exploited [6]. This and other GR main characteristics can be evaluated within PHDOM in a more "economic" way, using the effective-field method (see, e.g., Refs. [6,7,10]). The method has been introduced in RPA-based approaches by Migdal [13]. Let $V_{LM}(\vec{r}) = V_L(r)Y_{LM}(\vec{n})$ be the isoscalar s-p external field (probing operator). Using a convolution of this field with the effective p-h propagator to define the effective field, one gets the equation for the effective-field radial components $\tilde{V}_L(r, \omega)$ ($\omega$ is the excitation energy), which is simpler than the respective Bethe-Goldstone equation:

$$\tilde{V}_L(r, \omega) = V_L(r) + \frac{F(r)}{r^2} \int A_L(r, r', \omega)\tilde{V}_L(r', \omega)dr'. \tag{2}$$

Here, $(rr')^{-2}A_L(r,r',\omega)$ is the radial $L$-component of the "free" p-h propagator related to excitations in the neutral channels, $A_L = \sum_{\alpha=n,p} A_L^\alpha$, where indexes $n$ and $p$ are related to the neutron and proton subsystems, respectively. Rather cumbersome expressions for these components are given in Refs. [6,10]. These expressions contain the occupation numbers $n_\mu$, the s-p radial bound-state wave functions $r^{-1}\chi_\mu(r)$ and energies $\varepsilon_\mu$ with $\mu = n_{r,\mu}, j_\mu, l_\mu$ (($\mu$) $\equiv j_\mu, l_\mu$) being the set of bound-state quantum numbers, and the kinematic factors $t^L_{(\lambda)(\mu)} = \frac{1}{\sqrt{2L+1}}\langle(\lambda)\|Y_L\|(\mu)\rangle$. Also included are the Green functions $g_{(\lambda)}(r,r',\varepsilon = \varepsilon_\mu \pm \omega)$ of the single-particle radial Schrodinger equation, which contains the term $[-iW(\omega) + \mathcal{P}(\omega)]f_\mu f_{WS}(r)$, added to a mean field, with $W(\omega)$ and $\mathcal{P}(\omega)$ being the imaginary and real parts of the strength of the energy-averaged p-h self-energy term responsible for the spreading effect, and $f_{WS}(r)$ and $f_\mu$ are the Woods-Saxon function and its diagonal matrix element, respectively.

The effective field of Eq. (2) determines the main characteristics of ISGMPRs. In particular, the expression for the strength function $S_L(\omega)$, related to the above-given probing operator, is the following:

$$S_L(\omega) = -\frac{1}{\pi} Im\, P_L(\omega), \qquad (3)$$

where $P_L(\omega)$ is the respective polarizability,

$$P_L(\omega) = \int V_L(r) A_L(r,r',\omega) \tilde{V}_L(r',\omega) dr dr'. \qquad (4)$$

The results of strength-function calculations can be verified using the weakly model-dependent energy-weighted sum rule $EWSR_L = \int \omega S_L(\omega) d\omega$ [14]:

$$EWSR_L = \frac{1}{4\pi}\frac{\hbar^2}{2M} A \left\langle \left(\frac{dV_L(r)}{dr}\right)^2 + L(L+1)\left(\frac{V_L(r)}{r}\right)^2 \right\rangle. \qquad (5)$$

Here, the averaging <...> is performed over the nuclear density $n(r) = n^n(r) + n^p(r)$. In the next section, the strength functions $S_L(\omega)$, calculated for nuclei under consideration, are presented in terms of the relative energy-weighted strength functions (fractions of $EWSR_L$):

$$y_L(\omega) = \omega S_L(\omega)/EWSR_L, \qquad (6)$$

normalized by the condition $x_L = \int y_L(\omega) d\omega = 1$. (We omit the factor $(2L+1)$ in Eq. (5) in accordance with the definition of the strength functions of Eqs. (3), and (4).

The ISGMPR next characteristic is the projected transition density also related to the above-given probing operator [6,10]. The transition-density radial $L$-component, $r^{-2}\rho_{V_L}(r,\omega)$, can be expressed in terms of the respective effective field under the condition that the external-field radial part is a real quantity:

$$\frac{1}{r^2}\rho_{V_L}(r,\omega) = -\frac{1}{\pi}\text{Im}\tilde{V}_L(r,\omega)/(F(r)S_L^{1/2}(\omega)). \qquad (7)$$

The radial dependence of the $L$-component, $\rho_{V_L}(r, \omega = \omega_{L(peak)})$, taken at the peak energy of the respective ISGMPR strength function, exhibits specific behavior depending on the nature of this GR. Namely, this radial dependence exhibits nodeless, one-node, and two-node behavior for the main-tone, overtone, and second-order overtone GRs, respectively [10].

The choice of the external-field radial part also depends on the GR nature. Namely, for the main-tone GRs, the isoscalar giant quadrupole and octupole resonances (ISGQR and ISGOR, respectively), the radial part is taken as $V_L(r) = r^L$. For the GRs, which are overtones of the corresponding spurious states, the isoscalar giant monopole and dipole resonances (ISGMR and ISGDR, respectively), the radial part is taken as $V_L(r) = r^L(r^2 - \eta_L <r^2>)$. The $L = 0$ isoscalar spurious state is the nucleus ground state, while the $L = 1$ spurious state is the isoscalar dipole state associated with centre-of-mass motion. The adopted parameters $\eta_L$ are found from the condition of the absence of spurious-state excitation by the overtone external field. For the PHDOM-based description of ISGMR one gets: $\eta_{L=0}=1$ [7]. The method of evaluating $\eta_{L=1}$ and a detailed description of the spurious isoscalar dipole state are given in Ref. [10]. Among the overtones of real isoscalar GRs, the overtones of ISGMR (i.e., ISGMR2) and ISGQR (i.e., ISGQR2) have the lowest excitation energy. The radial part of the respective external field is taken as $V_L^{ov}(r) = r^2(r^2 - \eta_L^{ov} <r^2>)$. The adopted parameters $\eta_L^{ov}$ are found from the condition of minimal main-tone excitation by the overtone external field: $\int V_L^{ov}(r) \rho_{V_L}(r, \omega_{L(peak)}) dr = 0$.

A possibility to describe probabilities of GR direct one-nucleon decay belongs to specific features of PHDOM. These probabilities can be also expressed in terms of the effective field [7,10]. The strength function of direct one-nucleon decay of ISGMPR into the channel $\mu$, corresponding to population of one-hole $\mu^{-1}$ state in the subsystem $\alpha$ of the respective product nucleus, is determined by the squared amplitudes of a "direct + semi-direct" reaction induced by the external field $V_{LM}(\vec{r})$:

$$S_{L,\mu}^{\uparrow,\alpha}(\omega) = \sum_{(\lambda)} n_\mu^\alpha (t_{(\lambda)(\mu)}^L)^2 \left| \int \chi_{\varepsilon=\varepsilon_\mu+\omega,(\lambda)}^{\alpha *}(r) \tilde{V}_L(r,\omega) \chi_\mu^\alpha(r) dr \right|^2. \qquad (8)$$

Here, $r^{-1}\chi_{\varepsilon>0,(\lambda)}^\alpha(r)$ is the radial one-nucleon continuum-state wave function, having the standing-wave asymptotic behavior. Being normalized to the $\delta$-function of the energy in the $W = \mathcal{P} = 0$ limit, this wave function obeys the mentioned Schrodinger equation, in which the above-described complex term is added to the mean field. The partial branching ratio of direct one-nucleon decay of the ISGMPR into the channel $\mu$, $b_{L,\mu}^{\uparrow,\alpha}$, is determined by the strength functions of Eqs. (8), (3), and (4):

$$b_{L,\mu}^{\uparrow,\alpha}(\omega_{12}) = \int_{\omega_{12}} S_{L,\mu}^{\uparrow,\alpha}(\omega)d\omega \,/\, \int_{\omega_{12}} S_L(\omega)d\omega \,. \tag{9}$$

Here, $\omega_{12} = \omega_1 - \omega_2$ is an energy interval, that includes the considered GR. The total branching ratios $b_{L,tot}^{\uparrow,\alpha} = \sum_\mu b_{L,\mu}^{\uparrow,\alpha}$ determine the branching ratio of statistical (mainly neutron) decay: $b_L^{\downarrow} = 1 - \sum_{\alpha=n,p} b_{L,tot}^{\uparrow,\alpha}$. Note that in the cRPA limit ($W = \mathcal{P} = 0$) $\sum_{\alpha=n,p} b_{L,tot}^{\uparrow,\alpha} = b_L^{\uparrow} = 1$ (unitary condition), and $b_L^{\downarrow} = 0$.

### III. CHARACTERISTICS OF ISGMPRs

As mentioned in the Introduction, the following input quantities are used in the PHDOM-based description of main characteristics of ISGMPRs: (i) a realistic phenomenological, partially self-consistent mean field (described in detail in Refs. [10,15]), (ii) the spin-less part of Landau-Migdal p-h interaction (Eq. (1)), and (iii) the imaginary part $W(\omega)$ of the strength of the energy-averaged p-h self-energy term responsible for the spreading effect.

(i) The mean field contains the central (Woods-Saxon-type) and spin-orbit isoscalar terms (with intensities $U_0$ and $U_{ls}$, respectively), the isovector $\frac{1}{2}\tau^{(3)}v(r)$ and Coulomb terms. The isoscalar terms contain also the Woods-Saxon size and diffuseness parameters ($r_0$ and $a$, respectively). The symmetry potential $v(r) = 2F'n^{(-)}(r)$ and Coulomb term $U_C(r)$ are calculated self-consistently via the neutron-excess and proton density ($n^{(-)}(r) = n^n(r) - n^p(r)$ and $n^p(r)$, respectively). Due to the above-given isospin self-consistency condition, the strength parameter $F'$ of Eq. (1) might be related to mean-field parameters, which are found for doubly-closed-shell nuclei $^{48}$Ca, $^{132}$Sn, and $^{208}$Pb from the description of observable single-quasiparticle spectra in the respective even-odd and odd-even nuclei. The mean-field parameters are listed in Table I for $^{48}$Ca and $^{132}$Sn (for $^{208}$Pb, the parameters are given in Ref. [10]). Being based on these data, one gets the mean-field parameters for an arbitrary medium-heavy spherical nucleus by means of an interpolation procedure (see Appendix). The parameters obtained in such a way for $^{90}$Zr are also given in Table I.[1]

(ii) The isoscalar and isovector strengths of the spin-less part of the Landau-Migdal p-h interaction (Eq. (1)) are taken as $F(r) = Cf(r)$ and $F' = Cf'$, $C = 300$ MeV fm$^3$. The values of Landau-Migdal parameter $f'$, used below are given in Table I together with the mean-field parameters. The dimensionless strength $f(r)$ is parameterized in accordance with Ref. [13]:

$$f(r) = f^{ex} + (f^{in} - f^{ex})f_{WS}(r) . \qquad (10)$$

The small parameter $f^{in}$ is taken as an universal quantity, while the main parameter $f^{ex}$ in Eq. (10) is found for each considered nucleus (Table I) from the condition, that the energy of the spurious isoscalar dipole state is close to zero ($\cong 30\ keV$). The respective sum rule of Eq. (5) is well exhausted by the mentioned spurious state ($\cong 92\ \%$). The cRPA-based searching the spurious-state parameters is described in detail in Ref. [10].

---

[1] A search for the mean-field parameters was made in collaboration with V.I. Bondarenko.

(iii) In PHDOM implementations, the imaginary part of the strength of the energy-averaged p-h self-energy term is taken as a three-parametric function of the excitation energy:

$$2W(\omega) = \begin{cases} 0, & \omega < \Delta; \\ \alpha(\omega - \Delta)^2/[1 + (\omega - \Delta)^2/B^2], & \omega \geq \Delta. \end{cases} \quad (11)$$

Here, the adjustable ("spreading") parameters $\alpha$, $\Delta$ and $B$ can be called as the strength, gap, and saturation parameters, respectively. The use of the function $W(\omega)$ of Eq. (11) for evaluation of the strength of the self-energy term real part, $\mathcal{P}(\omega)$, via the proper dispersive relationship [4] leads to a rather cumbersome expression, which can be found in Ref. [16]. Below, we employ the values of "spreading" parameters found from the PHDOM-based description of the observable total width (full width at the half maximum (FWHM)) of $L = 0 - 3$ ISGMPRs in the $^{208}$Pb nucleus: $\alpha = 0.20$ MeV$^{-1}$, $\Delta = 3$ MeV, $B = 4.5$ MeV [10].

The first step in the PHDOM-based description of $L = 0 - 3$ ISGMPRs is calculation (according to Eqs. (2)-(4)) of the strength functions $S_L(\omega)$ related to the probing operators $V_L(\vec{r})$ given in Section II. After evaluating the $L = 0$, and 2 strength functions, one gets (according to the prescription given in Section II) the probing-operator parameters $\eta_L^{ov}$ (Table I), and the ISGMR2 and ISGQR2 strength functions. Calculated for nuclei under consideration, the above-listed strength functions are presented in terms of the energy-weighted-sum-rule fractions, $y_L(\omega)$ and $y_L^{ov}(\omega)$, of Eqs. (5), and (6). The $L = 0$ and $L = 2$ fractions together with the fractions of respective overtone GRs are shown in Figs. 1 and 2, respectively. The $L = 1$ and $L = 3$ fractions are presented in Fig. 3.

The strength functions calculated within PHDOM allow us to evaluate for a given excitation energy interval $\omega_{12} = \omega_1 - \omega_2$ the following ISGMPR parameters: (i) the integrated EWSR fraction (fraction parameter), $x$; (ii) the main-peak energy, $\omega_{peak}$; (iii) the centroid energy, $\bar{\omega}$, defined as the ratio of the first to zero moments of the respective strength function; and (iv) the total width (FWHM). The above-listed parameters evaluated for ISGMPRs in nuclei under consideration are given in Tables II-IV together with available experimental data.

Some of the evaluated GR parameters are compared in Tables II and III with respective results of calculations obtained within the self-consistent microscopic RPA based approach of Hartree-Fock (HFRPA) [11], using the SkT1[17] Skyrme effective nucleon-nucleon interaction. The SkT1 interaction is of the standard form with 10 parameters, which are determined by fitting results of HF calculations to experimental data on nuclear ground states properties of nuclei. We note that the SkT1 interaction is associated with an effective mass $m^* = 1$, which employed in the PHDOM. Detailed of the numerical method employed in the HFRPA calculations of the response functions and the centroid energies of giant resonances can be found in Refs. [11,18,19]. For accuracy, calculations of the response functions, of multipolarity $L = 0 - 3$, are carried in the $0 - 100$ MeV range of excitation energy. Also, for proper comparison with experimental data, a Lorentzian smearing of the calculated response functions is employed with values of smearing widths $\Gamma$ obtained experimentally (see Refs. [20,23]).

The next characteristic of ISGMPRs is the projected transition density of Eq. (7), $\rho_{V_L}(r, \omega)$. Being evaluated within PHDOM for nuclei under consideration, the radial (one-dimensional) transition densities taken at the peak energy of the respective GR are shown in Figs. 4 -6.

Turning to direct one-nucleon decay of ISGMPRs, we show in Tables V-VII the partial and total branching ratios, $b_{L,\mu}^{\uparrow,\alpha}$ and $b_{L,tot}^{\uparrow,\alpha}$, evaluated within PHDOM (in accordance with Eqs. (8), and (9)) for nuclei under consideration. For a comparison of the calculated partial branching ratio with the respective experimental value, when become available, it is reasonable to use the quantity $\check{b}_{L,\mu}^{\uparrow,\alpha} = (SF)_\mu b_{L,\mu}^{\uparrow,\alpha}$, where $(SF)_\mu$ is the spectroscopic factor of the product-nucleus one-hole state $\mu^{-1}$. In such a way, one can take into account a more complicated structure of the mentioned state.

The above-presented calculation results obtained within PHDOM for the main characteristics and parameters of ISGMPRs in medium-heavy closed-shell nuclei are similar to those obtained earlier for $^{208}$Pb [10]. Comments to these results are the following.

(i) The strength-function main-peak energies exhibit a smooth $A$-dependence (close to $A^{-1/3}$ for $L = 0, 2$, and 3) peculiar to the shape resonances (Table II - IV). In Fig. 7, the $A$-dependences of the calculated peak energies are shown for the $L = 0 - 3$ resonances (with inclusion of the results obtained for $^{208}$Pb [10]) in a comparison with respective experimental data.

(ii) Comparing with unity the total branching ratio of direct one-nucleon decay $b_L^\uparrow$ (or the branching ratio of statistical decay $b_L^\downarrow = 1 - b_L^\uparrow$) defined for a certain excitation-energy interval $\omega_{12}$ (Sect. II) allows one to estimate contribution of Landau damping + s-p continuum (or contribution of the spreading effect) to formation of the strength functions $S_L(\omega)$. As follows from the data given in Tables

V – VII (and in Table V of Ref. [10]), lowest in energy the main peak of $L = 2$ strength function is formed mainly due to the spreading effect ($b_{L=2}^{\downarrow} > 75\%$), while having larger energy the main peak of $L = 0$, 1, and 3 strength functions is formed, as a rule, mainly due to Landau damping + s-p continuum ($b_L^{\uparrow} > 50\%$). The last statement can be related also to $L = 0$, and 2 overtone-GR strength functions (Figs. 1, and 2).

(iii) The main peak of the considered $L = 0 - 3$ relative strength functions exhausts the most part of $EWSR_L$ (the respective fraction parameters $x_L(\omega_{12})$ are given in Tables V – VII). The rest is distributed in a large excitation-energy interval (Figs. 1 – 3). In particular, one can see a rather weak pigmy resonance in the energy dependence of $L = 2$ and $L = 3$ main-tone GR relative strength functions. These pigmy resonances are a trace of "free $2\hbar\omega$" and "free $3\hbar\omega$" p-h excitations, respectively, responsible for formation of the collective states associated with the main peak of the considered GRs. ("$1\hbar\omega$" means the intershell energy interval). Since the p-h interaction in the isoscalar channel is attractive (Table I), the pygmy resonance is placed at the high-energy tail of the respective GR. The asymmetry of the main peak in the energy dependence of $L = 0$ GR relative strength function (Fig. 1, and Fig. 2 in Ref. [10]) might be assigned to a manifestation of the pygmy resonance related to "free $2\hbar\omega$" p-h excitations. This statement is supported by a comparison of the relative strength functions calculated within PHDOM and cRPA for ISGMR in $^{208}$Pb [6]. Low- and high-energy components of the $L = 1$, and 3 relative strength functions (Fig. 3) appear due to contribution of "free $1\hbar\omega$" and "free $3\hbar\omega$" p-h excitations in formation of these strength functions. The mentioned low-energy components are usually associated with soft isoscalar modes (see, e.g., Ref. [27]). Evaluated within PHDOM parameters of these components are also given in Tables II-IV.

(iv) The strength functions $S_L(\omega)$ evaluated within PHDOM for $L = 0 - 2$ GRs in $^{90}$Zr are compared with the respective strength functions deduced from an analysis of respective reaction cross sections of GR excitation [23] (Fig. 8). Similar comparison is shown in Fig. 9 for $EWSR_{L=0}$ fraction related to ISGMR in $^{90}$Zr [28]. Bearing in mind the use of the collective-model (energy-independent) transition density in the above-mentioned analysis, the agreement of calculated and experimental strength functions looks satisfactory.

(v) The ISGMPR parameters evaluated with the use of the calculated strength functions are, as a rule, in an acceptable agreement with available experimental data (Tables II – IV). Evaluated within PHDOM the GR centroid and peak energies of the ISGMPRs are found to be lower by about $2 - 13$ %, than the energies obtained within the HFRPA method of Ref. [11] (Tables II, and III). In connection with this method we note that PHDOM is not fully self-consistent model and, therefore, cannot be used for a description of bulk properties of nuclei.

(vi) As expected, the projected transition density (taken at the resonance peak-energy) exhibits a node-less radial dependence for $L = 2$, and 3 main-tone GRs (Figs. 5, and 6), a one-node dependence for $L = 0$, and 1 first-order overtone GRs (Figs. 4, and 6), a two-node dependence for the $L = 0$ second-order overtone GR (Fig. 4).

(vii) As expected, the total one-neutron decay branching ratio is markedly larger, than the total branching ratio for one-proton decay of considered ISGMPRs (Tables V-VII). Experimental data on partial probabilities of ISGMPRs direct one-nucleon decay are scant. Here, we note Ref. [29], where the branching ratio $b_{L=1}^{\uparrow,n} = \sum_\mu b_{L=1,\mu}^{\uparrow,n}$ is deduced from a common analysis of $^{90}$Zr($\alpha,\alpha'$)– AND $^{90}$Zr($\alpha,\alpha'$n)– reaction cross-sections. (The sum is taken over four valence neutron-hole states $\mu^{-1}$ in $^{89}$Zr, Table VI). In the analysis of Ref. [29] different excitation-energy intervals are used in the definition of the branching ratio: $\omega_{12} = 23 - 28$ MeV in numerator, and $\omega_{12} = 10 - 37$ MeV in denominator of Eq. (9). Using this definition and the values of spectroscopic factors given in Ref. [29], we get, in accordance with Eq. (9), the value $\check{b}_{L=1}^{\uparrow,n}$=11 %, which is approximately twice the respective experimental value 4.8(9) % [29]. The above-described specification of calculating within PHDOM the one-nucleon direct-decay branching ratios is used for ISGDR in $^{208}$Pb. Obtained in Ref. [10] the value $b_{L=1}^{\uparrow,n} = 29.8$ % (the sum is taken over five valance neutron-hole states) is reduced up to the value $\check{b}_{L=1}^{\uparrow,n}$=9.4 %, which is in agreement with the respective experimental value 10.5(16) % [29].

## IV. CONCLUDING REMARKS

In this work, we presented results of the calculations of main characteristics and parameters of six Isoscalar Giant Multipole Resonances in medium-mass closed-shell nuclei $^{48}$Ca, $^{90}$Zr, and $^{132}$Sn. The calculations are performed within the semi-microscopic Particle-Hole Dispersive Optical Model, in which main relaxation modes of (p-h)-type states associated with giant resonances are together taken into account. Namely this point makes possible to get, within a model, a rather full description of the main giant-resonance characteristics for a wide excitation-energy interval and, therefore, to get information about giant-resonance structure and decay mechanisms. In calculations, we employed a realistic, phenomenological mean field with parameters taken from independent data, and the spin-less part of Landau-Migdal p-h interaction. Parameters of this interaction are related to the mean field due to isospin symmetry and translation invariance of the model Hamiltonian. The phenomenological parameters of the p-h self-energy term responsible for the spreading effect are taken from our previous study of ISGMPRs in $^{208}$Pb. Thus, in the present study, no specific adjusted model parameters are used. The energy-averaged strength functions and projected (one-body) transition density are evaluated for all considered GRs. The partial strength functions for direct one-nucleon decay are also evaluated. The above-mentioned strength functions are further used to estimate the main GR parameters such as the peak- and centroid-energies, total width, and probabilities of direct one-nucleon decay. As a rule, the calculation results are in an acceptable agreement with available experimental data and can serve as predictions in cases where experimental data are not available. This study and the above-mentioned description of ISGMPRs in $^{208}$Pb support the statement that the used model is a powerful tool for theoretical studies of various giant resonances in closed-shell nuclei. Extension of the model to taking into account nucleon pairing in medium-heavy open-shell spherical nuclei is in progress.

**AKNOWLEDGEMENTS**

The authors are grateful to Prof. U. Garg for providing experimental data and discussion. This work is partially supported by the Russian Foundation for Basic Research, under Grant no. 19-02-00660 (M.L.G., B.A.T., M.H.U.), by the US Department of Energy, under Grant no. DE-FG03-93ER40773 (S.S.), and by the Program "Priority – 2030" for National Research Nuclear University "MEPHI" (M.H.U.), and Project No. FSWU-2020-0035 of the Ministry of Science and Higher Education of the Russian Federation (M.H.U.).

**APPENDIX: INTERPOLATION PROCEDURE**

Searching for mean-field parameters for medium-heavy spherical nuclei (48<$A$<208) is based on the data obtained for closed-shell nuclei $^{48}$Ca, $^{132}$Sn (Table I), and $^{208}$Pb [10]. The following parabolic-type interpolation procedure is proposed. Let $x(A)$ be one from mean-field parameters ($U_0, U_{ls}, a, f'$):

$$x(A) = \sum_{k=0,1,2} x_k (A - A_0)^k . \qquad (A1)$$

Here, $x_k$ are adjustable parameters, $A_0$ is a reference value (taken equal to 180). The parameters can be found from (A1) after the use of this equation for the basic values of the $A = 48, 132,$ and $208$ nuclei.

TABLE I. The mean-field and related model parameters used in PHDOM-based calculations of characteristics of ISGMPRs in nuclei under consideration. (Notations are given in the text). The values $r_0 = 1.21$ fm and $f^{in} = 0.0875$ are taken as universal quantities.

| Nucleus | $U_0$, MeV | $U_{ls}$, MeV fm$^2$ | $a$, fm | $f'$ | $-f^{ex}$ | $\eta^{ov}_{L=0}$ | $\eta^{ov}_{L=2}$ |
|---|---|---|---|---|---|---|---|
| $^{48}$Ca | 54.34 | 32.09 | 0.576 | 1.13 | 2.556 | 3.07 | 2.00 |
| $^{90}$Zr | 55.06 | 34.93 | 0.612 | 1.05 | 2.580 | 3.17 | 1.77 |
| $^{132}$Sn | 55.53 | 35.98 | 0.633 | 0.999 | 2.536 | 2.72 | 1.85 |

**TABLE II.** The parameters of ISGMPRs in $^{48}$Ca evaluated within PHDOM and presented in a comparison with available experimental data and results of HFRPA calculations using Skyrme interaction [11]. (Notations are given in the text).

| L | $\omega_1 - \omega_2$, MeV | $x_L$, % | $\bar{\omega}_L$, MeV | $\omega_{L(peak)}$, MeV | $\Gamma_{L(FWHM)}$, MeV | |
|---|---|---|---|---|---|---|
| 0 | 5 – 35 | 99 | 19.7 | 19.5 | 8.5 | PHDOM |
| | | | 20.22 | 20.65 | | HFRPA |
| | 9.8 – 40.2 | $95^{+11}_{-15}$ | $19.88^{+0.14}_{-0.18}$ | - | $6.68^{+0.31}_{-0.36}$ | expt. [20] |
| | 10 – 31 | $78^{+4}_{-3}$ | 19.5±0.1 | | | expt. [21] |
| | 9.5 – 25.5 | | 18.40±0.13 | | | expt. [22] |
| 0 ov | 5 – 25 | 26 | 18.0 | 15.3; 23.6 | | PHDOM |
| | 25 – 50 | 63 | 35.3 | 26.9; 33.5 | | PHDOM |
| 1 (LE) | 4 – 16 | 12 | 11.2 | 10.0 | 2.1 | PHDOM |
| | | $20^{+12}_{-8}$ | | $16.69^{+0.19}_{-0.13}$- | $6.24^{+1.49}_{-0.11}$ | expt. [20] |
| 1 (HE) | 16 – 50 | 87 | 28.7 | 30.4 | 17.8 | PHDOM |
| | | | 30.13 | 32.70 | | HFRPA |
| | | $160^{+90}_{-50}$ | | $37.28^{+0.71}_{-1.98}$ | $14.95^{+3.49}_{-0.11}$ | expt. [20] |
| 2 | 5 – 35 | 97 | 15.9 | 15.9 | 2.4 | PHDOM |
| | | | 17.94 | 17.77 | | HFRPA |
| 2 ov | 8 – 24 | 32 | 17.4 | 19.8 | 8.1 | PHDOM |
| | 24 – 50 | 62 | 34.8 | 26.0 ; 33.9 | | PHDOM |
| 3 (LE) | 4 – 16 | 21 | 8.8 | 7.7 | 1.2 | PHDOM |
| 3 (HE) | 16 – 50 | 72 | 28.7 | 28.5 | 4.6 | PHDOM |
| | | | 31.62 | 32.85 | | HFRPA |

**TABLE III.** The same as in Table II, but for $^{90}$Zr.

| L | $\omega_1 - \omega_2$, MeV | $x_L$, % | $\bar{\omega}_L$, MeV | $\omega_{L(\text{peak})}$, MeV | $\Gamma_{L(\text{FWHM})}$, MeV | |
|---|---|---|---|---|---|---|
| 0 | 8 – 30 | 99 | 17.7 | 16.2 | 4.0 | PHDOM |
| | | | 18.01 | 17.51 | | HFRPA |
| | | 84 | 17.9±0.1 | 17.1 | 4.4 | expt. [23] |
| | | 95±6 | 18.13±0.09 | 16.55±0.08 | 4.2±0.3 | expt. [24] |
| | | | | 16.76±0.12 | 4.96±0.32 | expt. [25] |
| 0 ov | 8 – 25 | 31 | 19.3 | 19.6 | 5.9 | PHDOM |
| | 25 – 50 | 63 | 34.3 | 30.5 | | PHDOM |
| 1 (LE) | 5 – 15 | 11 | 11.1 | 12.20 | 2.5 | PHDOM |
| | | 9.2±2.1 | | 17.5±0.2 | 5.4±0.7 | expt. [23] |
| | | 7.9±2.9 | | 17.8±0.5 | 3.7±1.2 | expt. [26] |
| 1 (HE) | 15 – 40 | 87 | 26.7 | 29.7 | 11.3 | PHDOM |
| | | | 29.77 | 30.29 | | HFRPA |
| | | 49±6 | | 27.4±0.5 | 10.1±2.0 | expt. [23] |
| | | 67±8 | | 26.9±0.7 | 12.0±1.5 | expt. [26] |
| | | | | 27.76±0.88 | 11.28±2.5 | expt. [25] |
| | | 92±21 | | | | |
| | | | 25.5 | 0.5 | | |
| 2 | 5 – 35 | 96 | 13.8 | 13.5 | 2.7 | PHDOM |
| | | | 15.25 | 15.03 | | HFRPA |
| | | | | 14.56±0.20 | 4.94±0.20 | expt. [23] |
| | | | | 13.99±0.07 | 7.44±.030 | expt. [25] |
| 2 ov | 8 – 25 | 36 | 18.2 | 16.8 | 3.6 | PHDOM |
| | 25 – 50 | 61 | 34.5 | 38.0 | | PHDOM |
| 3 (LE) | 4 – 15 | 25 | 6.9 | 5.3 ; 11.6 | | PHDOM |
| 3 (HE) | 15 – 40 | 67 | 25.1 | 24.6 | 3.8 | PHDOM |
| | | | 28.21 | 27.72 | | HFRPA |

**TABLE IV.** The parameters of ISGMPRs in $^{132}$Sn evaluated within PHDOM.

| L | $\omega_1 - \omega_2$, MeV | $x_L$, % | $\bar{\omega}_L$, MeV | $\omega_{L(peak)}$, MeV | $\Gamma_{L(FWHM)}$, MeV |
|---|---|---|---|---|---|
| 0 | 5 – 30 | 100 | 15.6 | 15.4 | 4.7 |
| 0 ov | 5 – 25 | 46 | 15.5 | 11.7 ; 19.3 | |
| | 25 – 50 | 52 | 34.1 | 34.9 | |
| 1 (LE) | 5 – 15 | 18 | 9.7 | 6.8 ; 10.4 | |
| 1 (HE) | 15 – 40 | 83 | 24.8 | 26.0 | 15.7 |
| 2 | 4 – 25 | 88 | 12.0 | 12.0 | 2.7 |
| 2 ov | 4 – 20 | 37 | 13.1 | 13.4 | 7.0 |
| | 20 – 40 | 55 | 28.9 | 23.9 ; 33.7 | |
| 3 (LE) | 3 – 15 | 35 | 5.1 | | |
| 3 (HE) | 15 – 40 | 65 | 22.6 | 21.9 | 3.3 |

**TABLE V.** The partial and total branching ratios for direct one-nucleon decay of the ISGMPR into the channel $\mu$. The evaluated within PHDOM branching ratios (in %) for $^{48}$Ca are given with indication of the respective excitation-energy intervals $\omega_{12}$ and fraction parameters $x_L$ (see text).

| $\mu^{-1}$ | $\varepsilon_\mu$, MeV | $b^\uparrow_{L=0,\mu}$ | $b^\uparrow_{L=1,\mu}$ | $b^\uparrow_{L=2,\mu}$ | $b^\uparrow_{L=3,\mu}$ |
|---|---|---|---|---|---|
| Neutron | | | | | |
| $1f_{7/2}$ | -7.95 | 34.3 | 31.2 | 20.5 | 49.1 |
| $1d_{3/2}$ | -13.84 | 16.3 | 12.0 | 3.0 | 4.1 |
| $2s_{1/2}$ | -14.95 | 11.4 | 9.0 | 3.5 | 5.0 |
| $1d_{5/2}$ | -18.37 | 15.6 | 14.8 | 0.04 | 16.5 |
| $b^{\uparrow,n}_{L,tot}$ | | 77.6 | 75.1 | 27.0 | 75.5 |
| Proton | | | | | |
| $2s_{1/2}$ | -14.92 | 6.4 | 6.3 | 0.04 | 4.9 |
| $1d_{3/2}$ | -15.96 | 3.8 | 5.7 | 0.004 | 1.4 |
| $1d_{5/2}$ | -20.56 | 1.2 | 6.6 | - | 2.0 |
| $b^{\uparrow,p}_{L,tot}$ | | 11.4 | 23.3 | 0.04 | 8.3 |
| $\omega_1 - \omega_2$, MeV | | 13 – 26 | 17 – 38 | 13 – 19 | 24 – 33 |
| $x_L$, % | | 87 | 76 | 65 | 46 |

**TABLE VI.** The same as in Table V, but for $^{90}$Zr.

| $\mu^{-1}$ | $\varepsilon_\mu$, MeV | $b^\uparrow_{L=0,\mu}$ | $b^\uparrow_{L=1,\mu}$ | $b^\uparrow_{L=2,\mu}$ | $b^\uparrow_{L=3,\mu}$ |
|---|---|---|---|---|---|
| Neutron | | | | | |
| $1g_{9/2}$ | -11.26 | 16.2 | 17.2 | 3.8 | 18.1 |
| $2p_{1/2}$ | -14.10 | 10.7 | 4.6 | 0.54 | 2.9 |
| $1f_{5/2}$ | -15.26 | 7.3 | 8.6 | 0.03 | 2.0 |
| $2p_{3/2}$ | -15.97 | 16.7 | 9.9 | - | 9.0 |
| $1f_{7/2}$ | -20.10 | 0.01 | 11.2 | - | 3.0 |
| $2s_{1/2}$ | -25.27 | | 4.0 | - | 0.17 |
| $b^{\uparrow,n}_{L,tot}$ | | 50.9 | 61.0 | 4.4 | 35.2 |
| Proton | | | | | |
| $2p_{1/2}$ | -6.91 | 7.7 | 4.2 | 0.52 | 3.9 |
| $2p_{3/2}$ | -8.70 | 12.7 | 9.0 | 0.20 | 9.4 |
| $1f_{5/2}$ | -9.89 | 1.2 | 5.6 | 0.01 | 1.7 |
| $1f_{7/2}$ | -14.84 | | 6.2 | - | 2.0 |
| $2s_{1/2}$ | -18.17 | | 3.5 | - | 0.65 |
| $b^{\uparrow,p}_{L,tot}$ | | 21.6 | 28.5 | 0.7 | 17.7 |
| $\omega_1 - \omega_2$, MeV | | 14 – 21 | 20 – 33 | 11 – 16 | 21 – 28 |
| $x_L$, % | | 79 | 70 | 60 | 42 |

**TABLE VII.** The same as in Table V, but for $^{132}$Sn.

| $\mu^{-1}$ | $\varepsilon_\mu$, MeV | $b^\uparrow_{L=0,\mu}$ | $b^\uparrow_{L=1,\mu}$ | $b^\uparrow_{L=2,\mu}$ | $b^\uparrow_{L=3,\mu}$ |
|---|---|---|---|---|---|
| Neutron | | | | | |
| $1h_{11/2}$ | -7.11 | 13.4 | 15.6 | 6.0 | 14.8 |
| $2d_{3/2}$ | -7.32 | 13.1 | 5.2 | 4.8 | 5.3 |
| $3s_{1/2}$ | -7.91 | 5.2 | 2.5 | 4.5 | 1.6 |
| $1g_{7/2}$ | -9.51 | 10.1 | 8.7 | 0.89 | 2.8 |
| $2d_{5/2}$ | -9.98 | 17.7 | 9.6 | 7.9 | 13.1 |
| $1g_{9/2}$ | -14.97 | 7.9 | 12.0 | - | 4.3 |
| $2p_{1/2}$ | -16.61 | 0.82 | 3.9 | - | 1.2 |
| $2p_{3/2}$ | -18.02 | 0.34 | 8.1 | - | 10.2 |
| $1f_{5/2}$ | -18.93 | 0.12 | 4.6 | - | 0.59 |
| $b^{\uparrow,n}_{L,tot}$ | | 68.7 | 82.4 | 24.1 | 54.0 |
| Proton | | | | | |
| $1g_{9/2}$ | -15.17 | | 2.1 | - | 0.33 |
| $2p_{1/2}$ | -16.42 | | 1.9 | - | 0.01 |
| $2p_{3/2}$ | -17.79 | | 4.2 | - | 0.002 |
| $b^{\uparrow,p}_{L,tot}$ | | 0 | 8.2 | 0 | 0.34 |
| $\omega_1 - \omega_2$, MeV | | 10 – 21 | 18 – 31 | 9 – 15 | 19 – 24 |
| $x_L$, % | | 85 | 63 | 62 | 35 |

**FIG. 1.** The relative energy-weighted strength functions ($EWSR_L$ fractions) calculated within PHDOM for ISGMR (solid line) and ISGMR2 (thin line) in nuclei under consideration.

$y_{L=0}(\omega), y_{L=0}^{ov}(\omega)$ (MeV$^{-1}$)

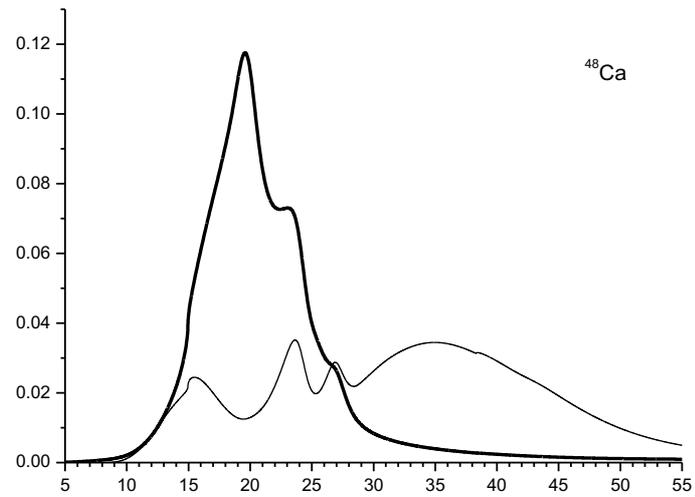

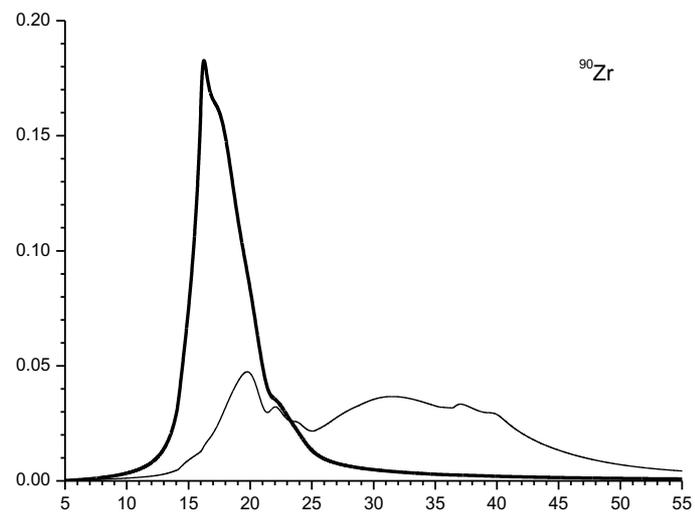

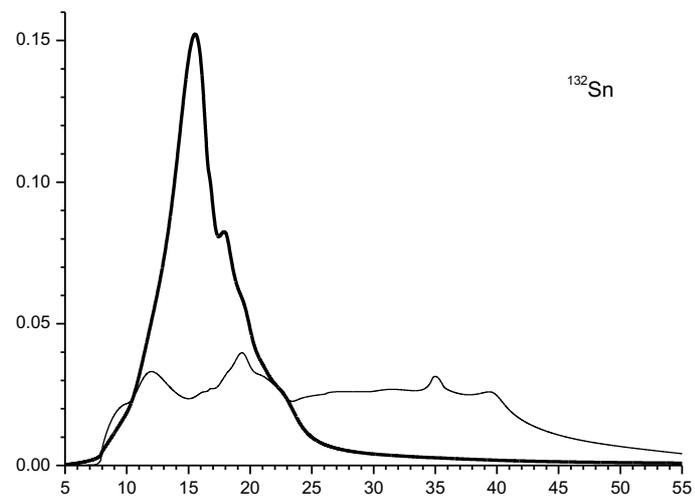

$\omega$ (MeV)

**FIG. 2.** The same as in Fig. 1, but for ISGQR and ISGQR2.

$y_{L=2}(\omega), y_{L=2}^{ov}(\omega)$ (MeV$^{-1}$)

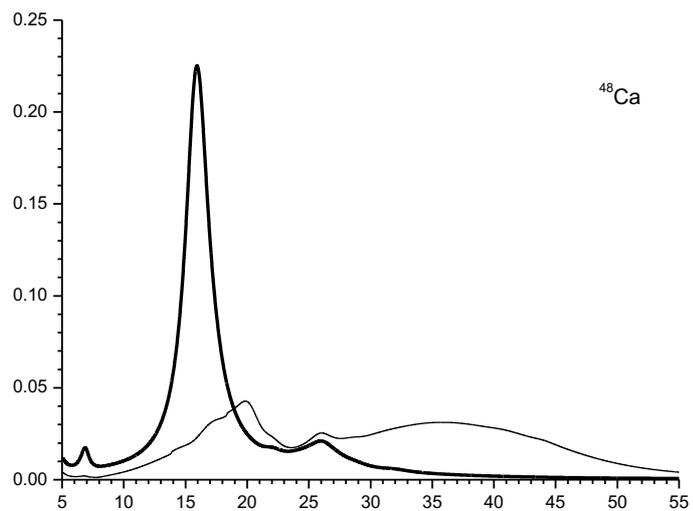

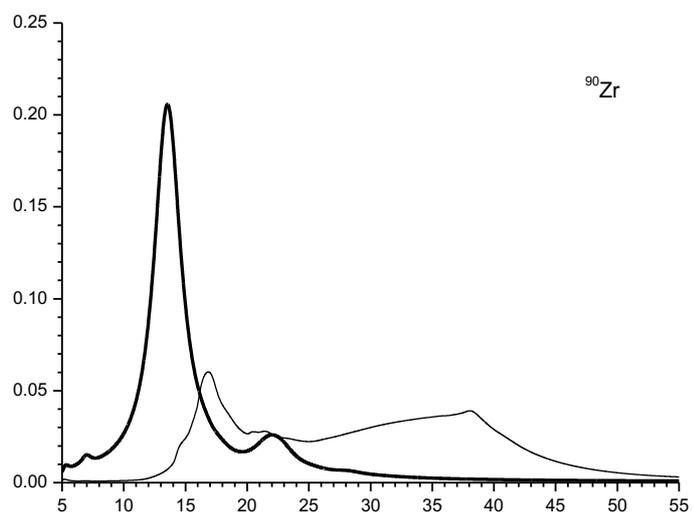

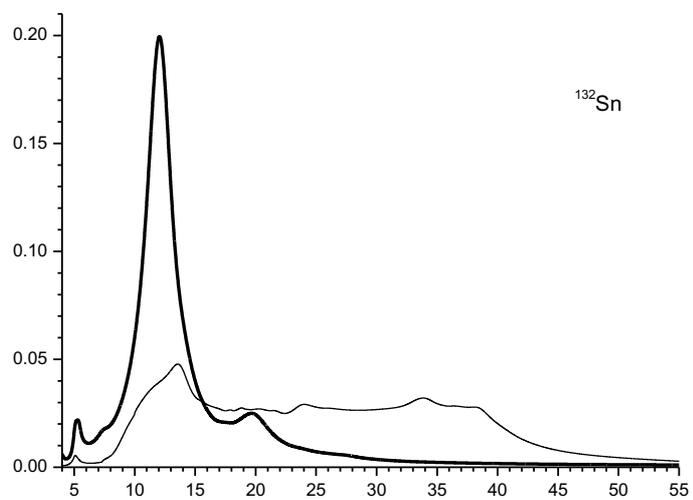

$\omega$ (MeV)

**FIG. 3**. The same as in Fig. 1, but for ISGDR and ISGOR.

$y_{L=1}(\omega), y_{L=3}(\omega)$ (MeV$^{-1}$)

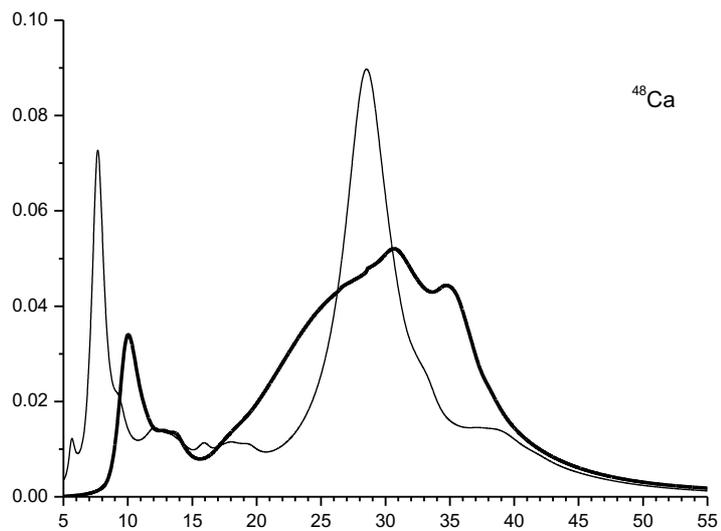

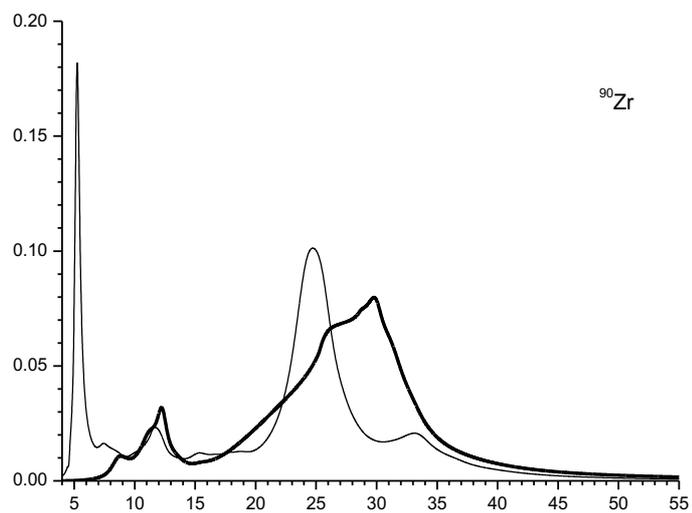

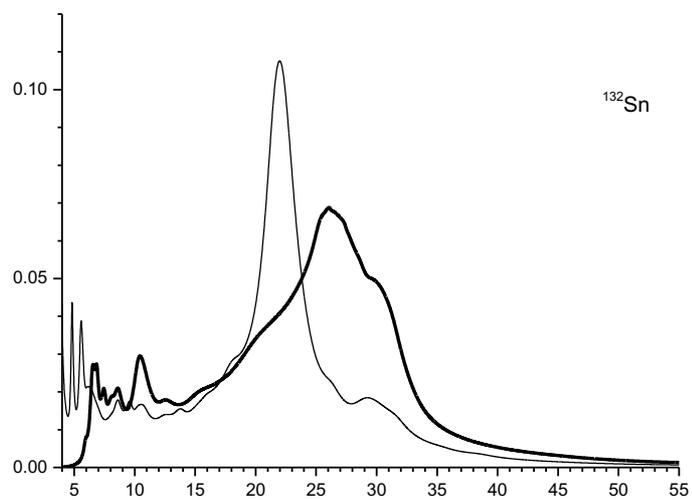

$\omega$ (MeV)

**FIG. 4.** The projected radial (one-dimensional) transition densities evaluated within PHDOM and taken at the peak energy of ISGMR (solid line) and ISGMR2 (thin line) in nuclei under consideration.

$\rho_{V_{L=0}}(r, \omega_{L=0(peak)}), \rho_{V_{L=0}^{ov}}(r, \omega_{L=0(peak)}^{ov})$ (fm$^{-1}$MeV$^{-1/2}$)

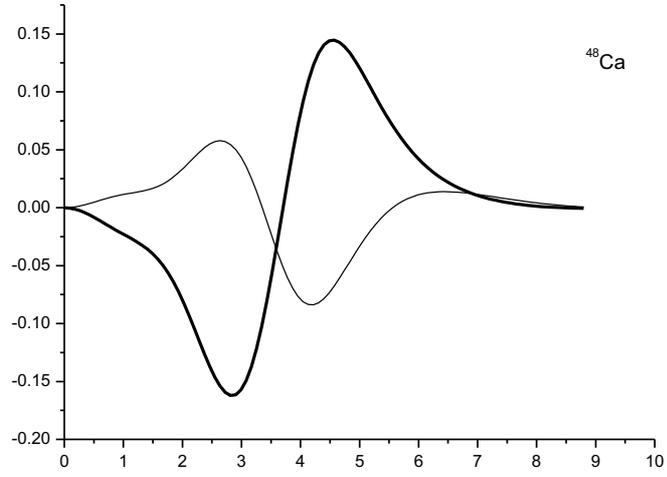

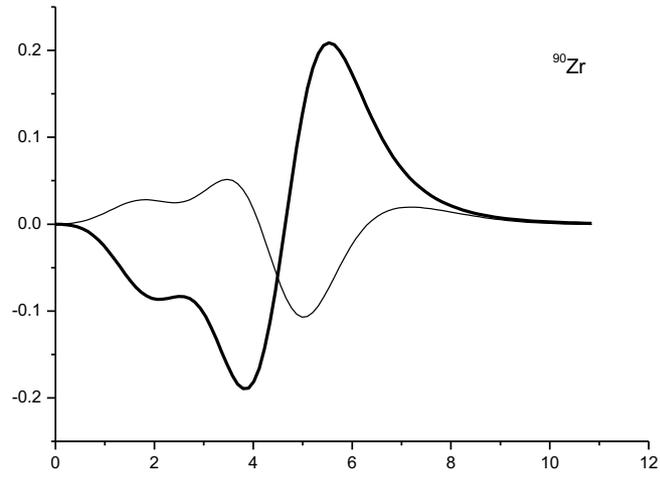

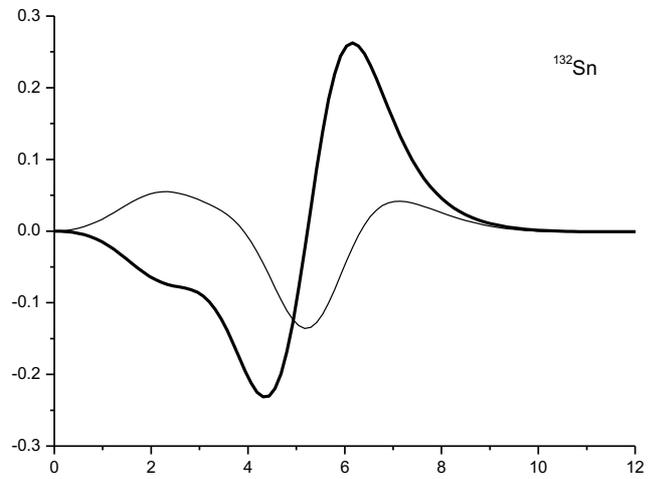

$r$ (fm)

**FIG. 5.** The same as in Fig. 4, but for ISGQR and ISGQR2.

$\rho_{V_{L=2}}(r, \omega_{L=2(peak)}), \rho_{V_{L=2}^{ov}}(r, \omega_{L=2(peak)}^{ov})$ (fm$^{-1}$MeV$^{-1/2}$)

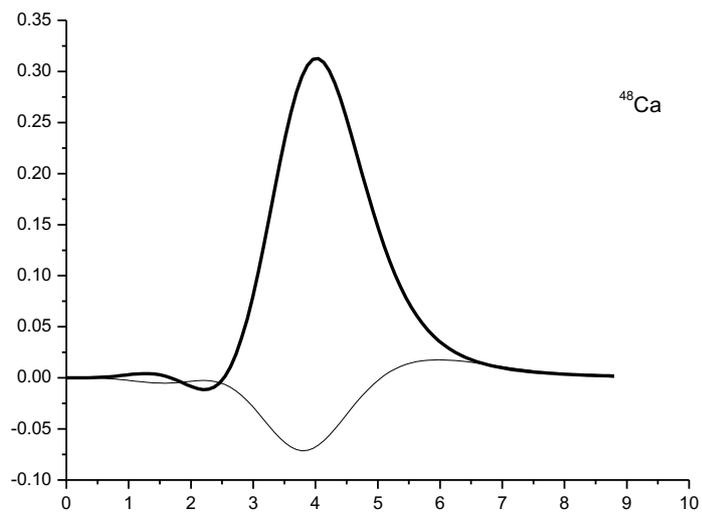

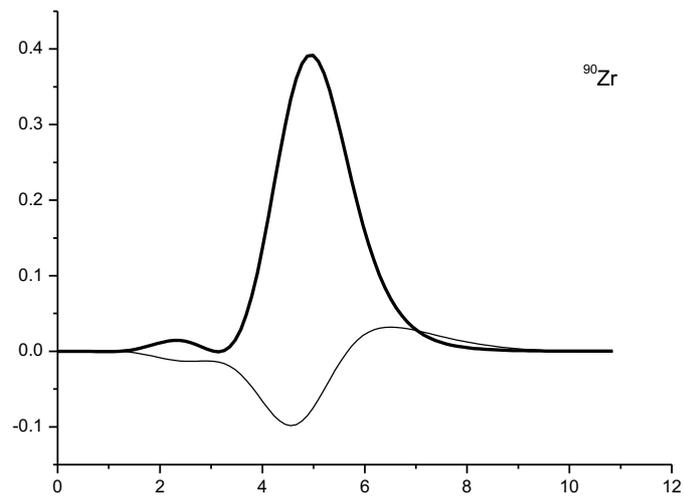

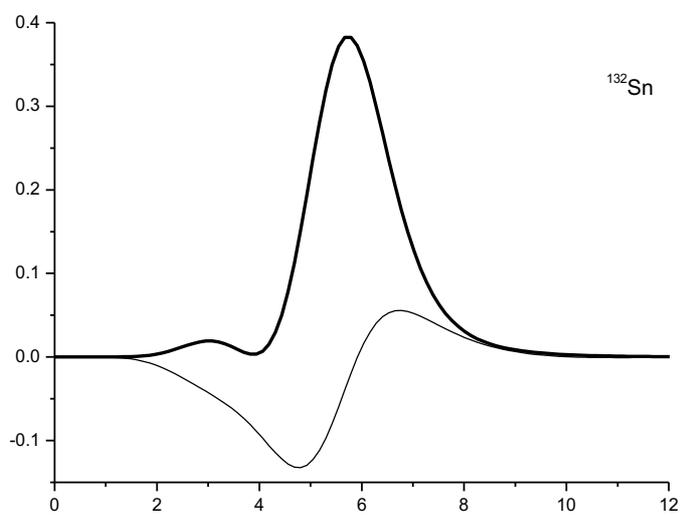

$r$ (fm)

**FIG. 6.** The same as in Fig. 4, but for ISGDR and ISGOR.

$\rho_{V_{L=1}}(r, \omega_{L=1(peak)})$, $\rho_{V_{L=3}}(r, \omega_{L=3(peak)})$ (fm$^{-1}$MeV$^{-1/2}$)

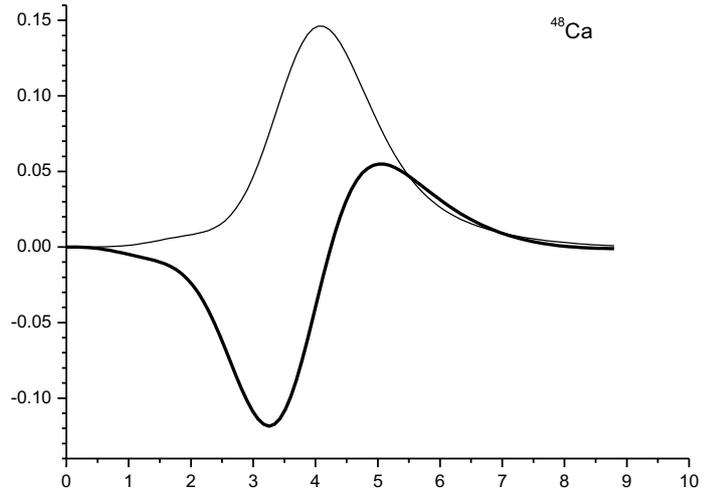

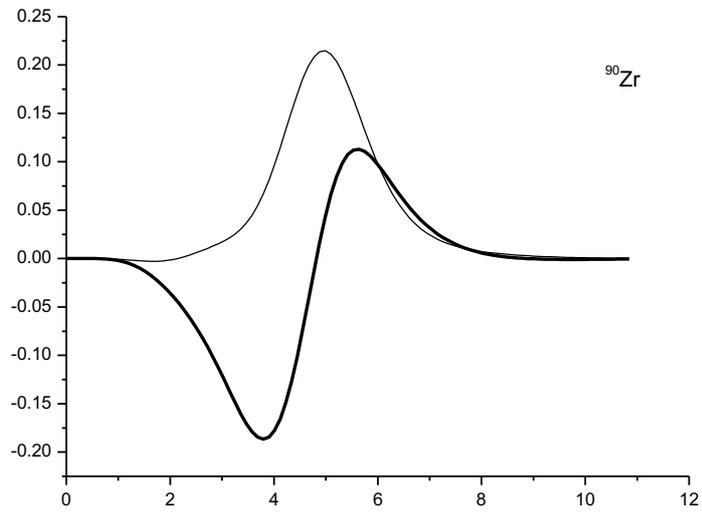

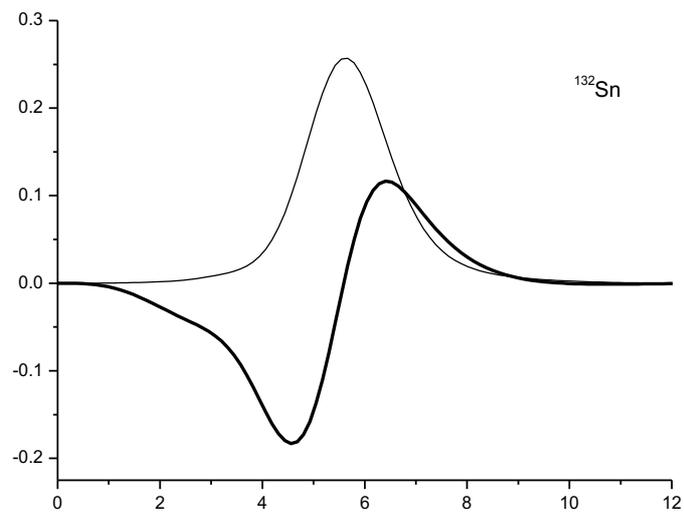

$r$ (fm)

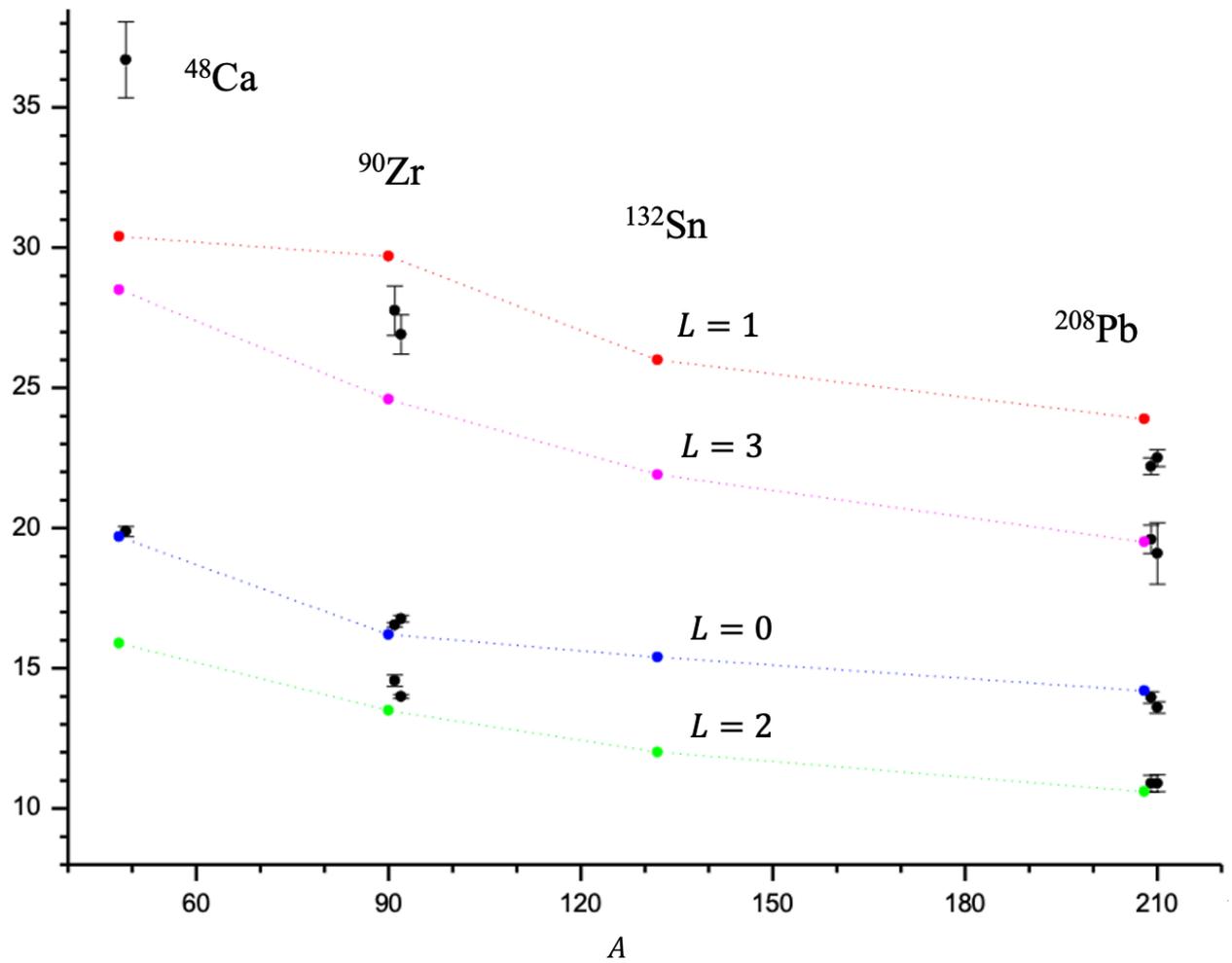

**FIG. 7.** Evaluated within PHDOM the peak energies of $L = 0 - 3$ ISGMPRs (the respective values for $^{208}$Pb [10] are included) in a comparison with respective experimental data.

**FIG. 8.** The strength functions $S_L(\omega)$ evaluated within PHDOM for $L = 0 - 2$ GRs in $^{90}$Zr in a comparison with respective strength functions (multiplied by proper normalization factors) deduced from an analysis of respective reaction cross sections of GR excitation [25].

$S_{L=0}(\omega)$ (fm$^4$ MeV$^{-1}$)

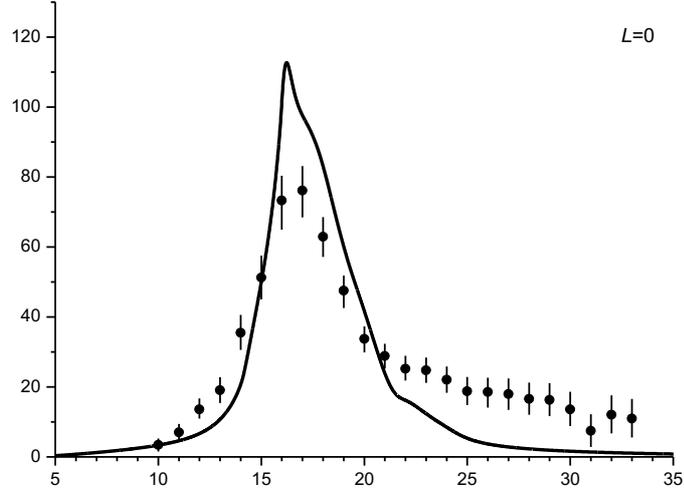

$S_{L=1}(\omega)$ (fm$^6$ MeV$^{-1}$)

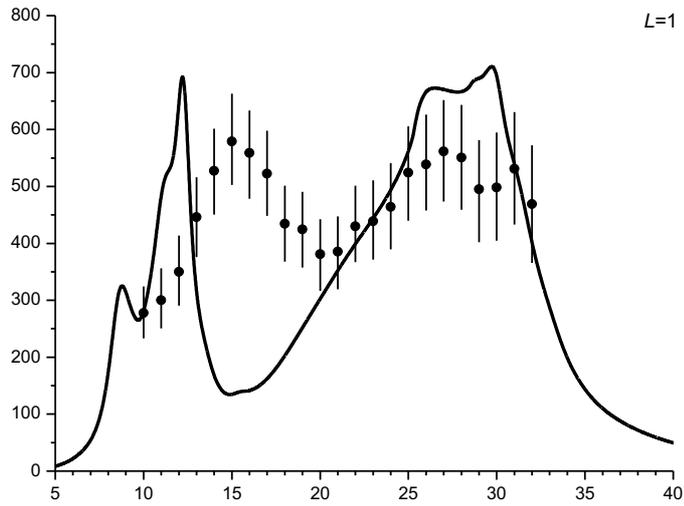

$S_{L=2}(\omega)$ (fm$^4$ MeV$^{-1}$)

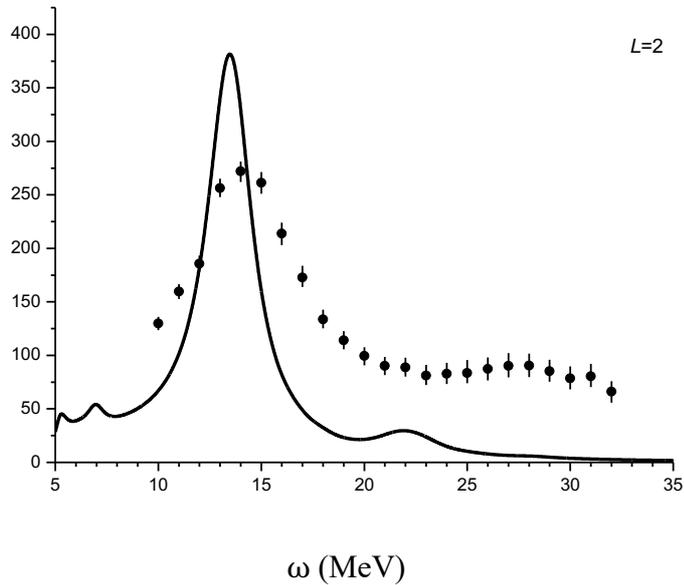

$\omega$ (MeV)

**FIG. 9.** The relative energy-weighted strength functions (the $EWSR_L$ fractions) calculated within PHDOM for ISGMR in $^{90}$Zr in a comparison with respective experimental data [28].

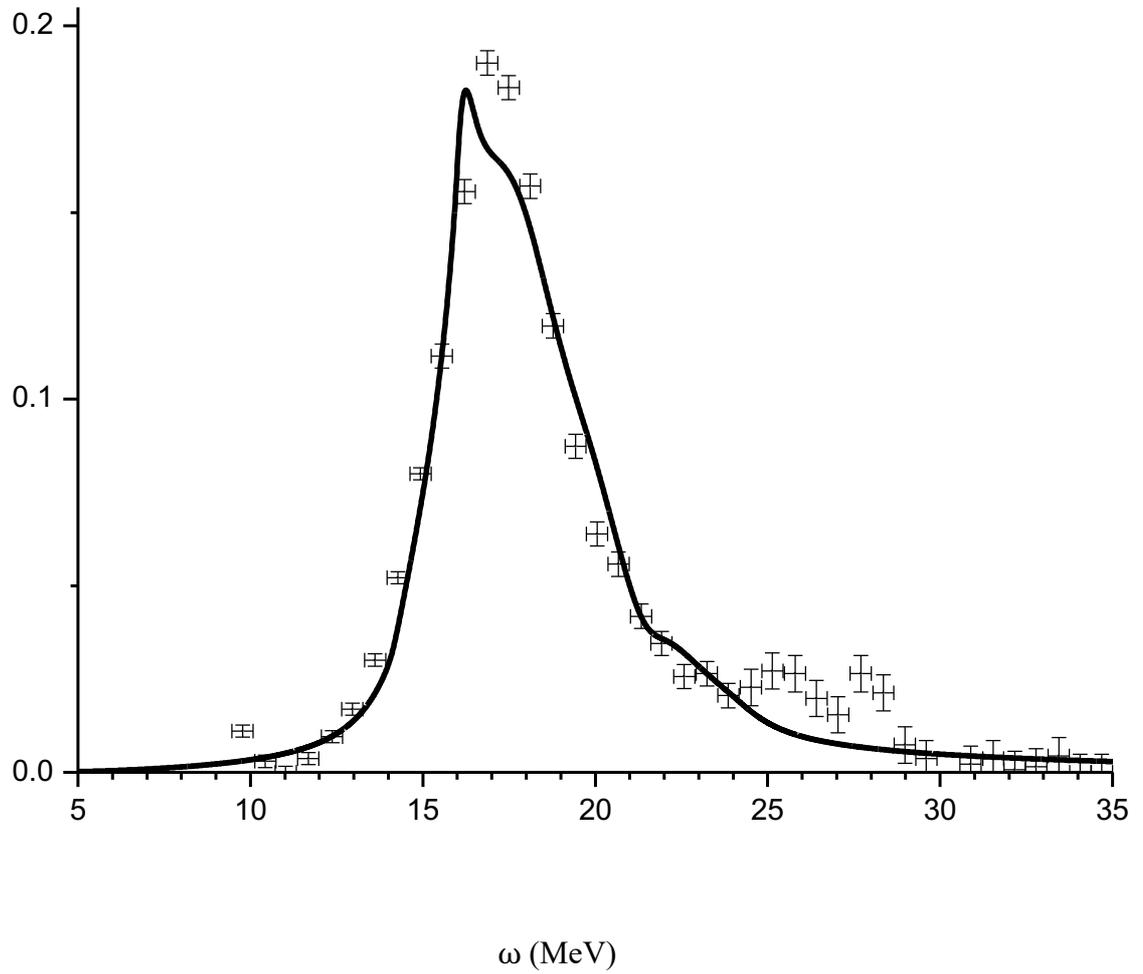

$y_{L=0}(\omega)$ (MeV$^{-1}$)

$\omega$ (MeV)